\documentclass[12pt]{article} 
\usepackage{graphicx}
\usepackage{epsfig} 
\usepackage{latexsym} 
\usepackage{latexsym}
\usepackage{amssymb} 
\usepackage{amsmath} 
\usepackage{cite}
\newcommand{\reef}[1]{(\ref{#1})}

\def\ie{{\it i.e.}}

\def\Tr{{\rm Tr}}

\def\zb{{\bar z}}

\def\bz{{\bar z}}

\def\half{\frac{1}{2}}

\DeclareSymbolFont{AMSb}{U}{msb}{m}{n}
\DeclareMathSymbol{\IN}{\mathbin}{AMSb}{"4E}
\DeclareMathSymbol{\IZ}{\mathbin}{AMSb}{"5A}
\DeclareMathSymbol{\IR}{\mathbin}{AMSb}{"52}
\DeclareMathSymbol{\Q}{\mathbin}{AMSb}{"51}
\DeclareMathSymbol{\II}{\mathbin}{AMSb}{"49}
\DeclareMathSymbol{\IC}{\mathbin}{AMSb}{"43}
\DeclareMathSymbol{\IP}{\mathbin}{AMSb}{"50}
\DeclareMathSymbol{\IH}{\mathbin}{AMSb}{"48}
\DeclareMathSymbol\IA{\mathalpha}{AMSb}{"41}
\DeclareMathSymbol\IS{\mathalpha}{AMSb}{"53}

\def\Q{{\cal Q}}

\begin{document}

\begin{flushright}
USC-04-04\\
DCPT--04/13                                       
\end{flushright}
\bigskip
\begin{center} {\Large \bf An Exact String Theory  Model of }

\bigskip\bigskip

{\Large \bf Closed Time--Like Curves and Cosmological Singularities }

\end{center}

\bigskip
\bigskip
\bigskip

\centerline{\bf Clifford V.
  Johnson${}^\natural$, Harald G. Svendsen${}^\sharp$}

\bigskip
\bigskip
\bigskip

  \centerline{\it ${}^\natural$Department of Physics and Astronomy }
\centerline{\it University of
Southern California}
\centerline{\it Los Angeles, CA 90089-0484, U.S.A.}
\centerline{\small \tt johnson1@usc.edu}
\bigskip
\bigskip

\bigskip
\bigskip

\centerline{\it ${}^\sharp$Centre for Particle Theory} \centerline{\it
  Department of Mathematical Sciences} \centerline{\it University of
  Durham} \centerline{\it Durham, DH1 3LE, U.K.}
\centerline{\small \tt h.g.svendsen@durham.ac.uk}


\bigskip
\bigskip


\begin{abstract}
  We study an exact model of string theory propagating in a
  space--time containing regions with closed time--like curves (CTCs)
  separated from a finite cosmological region bounded by a Big Bang
  and a Big Crunch. The model is an non--trivial embedding of the
  Taub--NUT geometry into heterotic string theory with a full
  conformal field theory (CFT) definition, discovered over a decade
  ago as a heterotic coset model. Having a CFT definition makes this
  an excellent laboratory for the study of the stringy fate of CTCs,
  the Taub cosmology, and the Milne/Misner--type chronology horizon
  which separates them. In an effort to uncover the role of stringy
  corrections to such geometries, we calculate the complete set of
  $\alpha^\prime$ corrections to the geometry. We observe that the key
  features of Taub--NUT persist in the exact theory, together with the
  emergence of a region of space with Euclidean signature bounded by
  time--like curvature singularities. Although such remarks are
  premature, their persistence in the exact geometry is suggestive
  that string theory theory is able to make physical sense of the
  Milne/Misner singularities and the CTCs, despite their pathological
  character in General Relativity.  This may also support the
  possibility that CTCs may be viable in some physical situations, and
  may be a natural ingredient in pre--Big--Bang cosmological
  scenarios.

\end{abstract}
\newpage \baselineskip=18pt \setcounter{footnote}{0}

\section{Introduction and Motivation}
\label{sec:introduction}

The Taub-NUT spacetime\cite{Taub:1951ez,Newman:1963yy} is an
interesting one.  We can write a metric for it as follows:
\begin{equation}
ds^2=-f_1(dt-l\cos\theta d\phi)^2+f_1^{-1}dr^2+(r^2+l^2)(d\theta^2+\sin^2\theta d\phi^2)\ ,
  \label{eq:taubnut}
\end{equation}
where 
\begin{equation}
f_1=1-2\frac{Mr+l^2}{r^2+l^2}\ .
  \label{eq:fone}
\end{equation}
The angles $\theta$ and $\phi$ are the standard angles parameterizing
an $S^2$ with ranges $0\leq \theta\leq\pi$, $0\leq \phi\leq 2\pi$. In
addition to simple time translation invariance, the metric has an
$SO(3)$ invariance acting as rotations on the $S^2$. To preserve
$d\xi=dt-l\cos\theta d\phi$, a time translation must also accompany a
general rotation. This makes $t$ periodic with period $4l\pi$, which
can be deduced by asking for there to be no conical singularities in
the North or South poles. The coordinate $t$ is fibred over the $S^2$
making a squashed $S^3$, and the full invariance  is under an $SU(2)$
action on this space.

There are two very different regions of this spacetime, as one moves
in $r$, distinguished by the sign of $f_1(r)$. The regions are
separated by the loci (with $S^3$ topology)
\begin{equation}
  \label{eq:horizons}
  r_\pm=M\pm\sqrt{M^2+l^2}\ ,
\end{equation}
where $f_1$ vanishes. They are, in a sense, horizons. The metric is
singular there, but there exist extensions the nature of which is
subtle in General Relativity (for a review, see ref.\cite{HawkingEllis}).
One of the things which we will discuss in detail later is the fact
that the string theory provides an extremely natural extension.

The region $r_-<r<r_+$ has $f_1(r)<0$. The coordinate $r$ plays the
role of time, and the geometry changes as a function of time. This is
the ``Taub'' cosmology, and spatial slices have the topology of an
$S^3$. The volume of the universe begins at $r=r_-$ at zero, it expands
to a maximum value, and then contracts to zero again at $r=r_+$. This
is a classical ``Big Bang'' followed by a classical ``Big Crunch''.

On either side of this Taub region, $f_{1}(r)>0$. The coordinate $t$
plays the role of time, and we have a static spatial geometry, but
since $t$ is periodic, it is threaded by closed time--like curves.
Constant radial slices have the topology of an $S^3$ where the time is
a circle fibred over the $S^2$. These regions are called the ``NUT'' regions.

It is fascinating to note that the Taub and NUT regions are connected.
There are geodesics which can pass from one region to another, and
analytic extensions of the metric can be written
down\cite{HawkingEllis}. The geometry is therefore interesting, since
it presents itself as a laboratory for the study of a cosmology which
naturally comes capped with regions containing CTCs. Classical physics
would seem to suggest that one can begin within the cosmological
region and after waiting a finite time, find that the universe
contained closed time--like loops.

It is an extremely natural question to ask whether or not this is an
artifact of classical physics, a failure of General Relativity to
protect itself from the apparent pathologies with which such time
machines seem to be afflicted. This leads to a closer examination of
the neighbourhood of the loci $f_1(r)=0$ located at $r=r_\pm$, which
we shall call (adopting common parlance) ``chronology horizons''. For
small $\tau=r-r_-$, we see that $f_1=-c\tau$, where $c$ is a constant,
and we get for the $(\tau,\xi)$ plane:
\begin{equation}
ds^2=-(c\tau)^{-1}d\tau^2+c\tau d\xi^2\ ,
  \label{eq:Misner}
\end{equation}
which is the metric of a two dimensional version of the ``Milne''
Universe, or ``Misner space''\cite{Misner}. It is fibred over the $S^2$.

There is an early study of cosmological singularities of this type in
a semi--classical quantum treatment, reported on in
ref.\cite{Hiscock:1982vq}. There, the vacuum stress--energy tensor for
a conformally coupled scalar field in the background is computed, and
it diverges at $\tau=0$. This is taken by some as an encouraging sign
that a full theory of quantum gravity might show that the geometry is
unstable to matter fluctuations and the appropriate back--reaction
should give a geometry which is modified at the boundaries between the
Taub and NUT regions. In fact, this is the basis of the ``chronology
protection conjecture'' of ref.\cite{Hawking:1992nk}, which suggests
(using Taub--NUT as a one of its key examples) that the full physics
will conspire to forbid the creation of CTCs in a spacetime that does
not already have them present, {\it i.e.,} the Misner geometry of the
chronology horizon is destroyed and replaced by a non--traversable
region\footnote{Even staying within Relativity, there are many who
  take an alternative view, by {\it e.g.,} showing that a
  non--divergent stress tensor can be obtained by computing in a
  different vacuum, thus calling into the question the need for such a
  conjecture. See for example,
  refs.\cite{Kim:1991mc,Thorne:1992gv,Li:1993ai,Li:1994xf,Tanaka:1995bh,Li:1996dba,Krasnikov:1996jn,Sushkov:1997hg,Visser:1997ek,Li:1998ka}
  and for a recent stringy example, see ref.\cite{Biswas:2003ku}.}.
The expectations of a full theory of quantum gravity in this regard
are (at least) two--fold: (1) It should prescribe exactly what types
of matter propagate in the geometry, and; (2) It should give a
prescription for exactly how the geometry is modified, incorporating
any back--reaction of the matter on the geometry in a self--consistent
way.

Since the papers of ref.\cite{Hiscock:1982vq,Hawking:1992nk}, a lot
has happened in fundamental physics. In particular, it is much clearer
that there is a quantum theory of gravity on the market. It should
allow us to study the questions above\footnote{Leaving aside the
  question of CTCs, cosmological singularities of Misner type have
  recently become relevant in the context of cosmologies inspired
  by string-- (and M--) theory. See for example
  ref.\cite{Khoury:2001bz}}.  Of course, we are
referring to string theory (including its not yet fully defined
non--perturbative completion in terms of M--theory).  While the theory
has yet to be developed to the point where we can address the physics
of spacetime backgrounds in as dextrous a way as possible, there are
many questions which we can ask of the theory, and in certain special
cases, we can study certain spacetime backgrounds in some detail.

In fact, as we will recall in the next section, the Taub--NUT
spacetime can be embedded into string theory in a way that allows its
most important features to be studied in a very controlled laboratory,
an {\it exact} conformal field theory\cite{Johnson:1994jw}. It is
therefore not just accessible as a solution to the leading order in an
expansion in small $\alpha^\prime$ (the inverse string tension), but
to all orders and beyond. Leading order captures only the physics of
the massless modes of the string, (the low energy limit) and so any
back--reaction effecting the geometry {\it via} high--energy effects
cannot be studied in this limit. With the full conformal field theory
one can in principle extract the complete geometry, including all the
effects of the infinite tower of massive string states that propagate
in it. We do this in the present paper and extract the fully corrected
geometry. We observe that the key features of the geometry {\it
  survive} to all orders in $\alpha^\prime$, even though placed in a
string theory setting without any special properties to forbid
corrections. This result means that a large family of high energy
effects which could have modified the geometry are survived by the
full string theory. The string seem to propagate in this apparently
pathological geometry with no trouble at all. It is of course possible
that the new geometry we find is unstable to the presence of a test
particle or string, but this type of effect does not show up in the
CFT in this computation. Such test--particle effects are important to
study\footnote{They have been found for the leading order geometry in its
  form as an orbifold of Minkowski space by a Lorentz
  boost\cite{Liu:2002ft,Liu:2002kb,Simon:2002ma,Lawrence:2002aj,Horowitz:2002mw}.}
in order to understand the complete fate of the geometry by studying
its stability against fluctuations. Our work here yields the fully
corrected geometry in which such probe computations should be carried
out in this context. More properly, the probe computation should be
done in the full conformal field theory, in order to allow the string
theory to respond fully to the perturbation.  The conformal field
theory discussed here is a complete laboratory for such studies, and
as it describes the Taub--NUT geometry, it provides the most natural
stringy analogue of this classic geometry within which to answer many
interesting questions\footnote{There are a number of other interesting
  conformal field theories (and studies thereof) which have been
  presented, which at low energy describe geometries which although
  are not Taub--NUT spacetimes, do share many of the key features in
  local patches. Some of them are listed in
  refs.\cite{Horava:1992am,Nappi:1992kv,Giveon:1992kb,Kiritsis:1994fd,Kounnas:1992wc,Tseytlin:1992xk,deVega:2001wq,Elitzur:2002rt,Cornalba:2002fi,Craps:2002ii,Buchel:2002kj,Fabinger:2002kr,Cornalba:2002nv,Cornalba:2003ze,Berkooz:2002je,Berkooz:2004re}.
  Refs.\cite{Berkooz:2002je,Berkooz:2004re} also contain useful
  comments and literature survey. There are also many papers on the
  properties of string theory in spacetimes with CTCs, such as the
  BMPV\cite{Breckenridge:1997is}
  spacetime\cite{Gibbons:1999uv,Dyson:2003zn,Jarv:2002wu,Fiol:2003yq,Maoz:2003yv,Brace:2003rk,Brace:2003st,Behrndt:2003gc,Drukker:2003mg,Brecher:2003wq,Brecher:2003rv,Hikida:2003yd,Bena:2004wt}
  and the G\"odel\cite{Godel:1949ga}
  spacetime\cite{Herdeiro:2002ft,Gimon:2003ms,Boyda:2002ba,Drukker:2003sc,Israel:2003cx,Drukker:2004zm,Gimon:2004if}.}.

In section~\ref{sec:stringytaubnut}, we recall the stringy Taub--NUT
metric discovered in ref.\cite{Johnson:1994jw}, and write it in a new
coordinate which gives it a natural extension exhibiting the Taub and
NUT regions and their connection {\it via} Misner space.  We also
recall the work of
refs.\cite{Johnson:1994ek,Kallosh:1994ba,Johnson:1994nj} which
demonstrates how to obtain the low energy metric as a stringy
embedding by starting with the standard Taub--NUT metric of
equation~\reef{eq:taubnut}. It is the ``throat'' or ``near--horizon''
region of this spacetime that was discovered in
ref.\cite{Johnson:1994jw}, where an exact conformal field theory (a
``heterotic coset model'') can be constructed which encodes the full
stringy corrections. We review the conformal field theory construction
in sections~\ref{sec:exact1} and~\ref{sec:exact2}, where the
Lagrangian definition is reviewed. Happily, the extension of the
throat geometry we present in section~\ref{sec:stringytaubnut}
(described by the same conformal field theory) contains all the
interesting features: the Taub region with its Big--Bang and
Big--Crunch cosmology, the NUT regions with their CTCs, and the Misner
space behaviour which separates them. Therefore we have a complete
string theory laboratory for the study of the properties of Taub--NUT,
allowing us to address many of the important questions raised in the
Relativity community. For example, questions about the analytic
extension from the NUT to the Taub regions are put to rest by the fact
that the full conformal field theory supplies a natural extension {\it
  via} the structure of $SL(2,\IR)$
(section~\ref{sec:stringytaubnut}). Further, having the full conformal
field theory means that we can construct the $\alpha^\prime$
corrections to the low energy metric, and we do so in
section~\ref{sec:exactgeometry}, capturing {\it all} of the
corrections, after constructing an exact effective action in
sections~\ref{sec:effective1} and~\ref{sec:effective2}.  We analyze
the exact metric in section~\ref{sec:properties}, and end with a
discussion in section~\ref{sec:discussion}, noting that there are many
questions that can be answered in this laboratory by direct
computation in the fully defined model.

\section{Stringy Taub--NUT}
\label{sec:stringytaubnut}

Taub--NUT spacetime, being an empty--space solution to the Einstein
equations, is trivially embedded into string theory with no further
work. It satisfies the low--energy equations of motion of any string
theory, where the dilaton is set to a constant and all the other
background fields are set to zero. This is not sufficient for what we
want to do, since we want to have a means of getting efficient
computational access to the stringy corrections to the geometry. A new
embedding must be found which allows such computational control.

This was achieved some time ago. An exact conformal field theory
describing the Taub--NUT spacetime (in a certain ``throat'' or
``near--horizon'' limit) was constructed in ref.\cite{Johnson:1994jw}.
This CFT will be described in the next section.  The geometry comes
with a non--trivial dilaton and anti--symmetric tensor field, together
with some electric and magnetic fields. The string theory is heterotic
string theory. This model is in fact the earliest non--trivial
embedding of Taub--NUT into string theory, and uses a novel
construction known as ``heterotic coset models'' in order to define
the theory\cite{Johnson:1994jw,
  Johnson:1995kv,Johnson:1994kh,Berglund:1996dv}. The technique was
discovered as a method of naturally defining $(0,2)$ conformal field
theories, {\it i.e.,} backgrounds particularly adapted to yielding
minimally supersymmetric vacua of the heterotic string. That aspect
will not be relevant here, since we will not tune the model in order
to achieve spacetime supersymmetry.

\begin{figure}[ht]
\begin{center}
\includegraphics[scale=0.5]{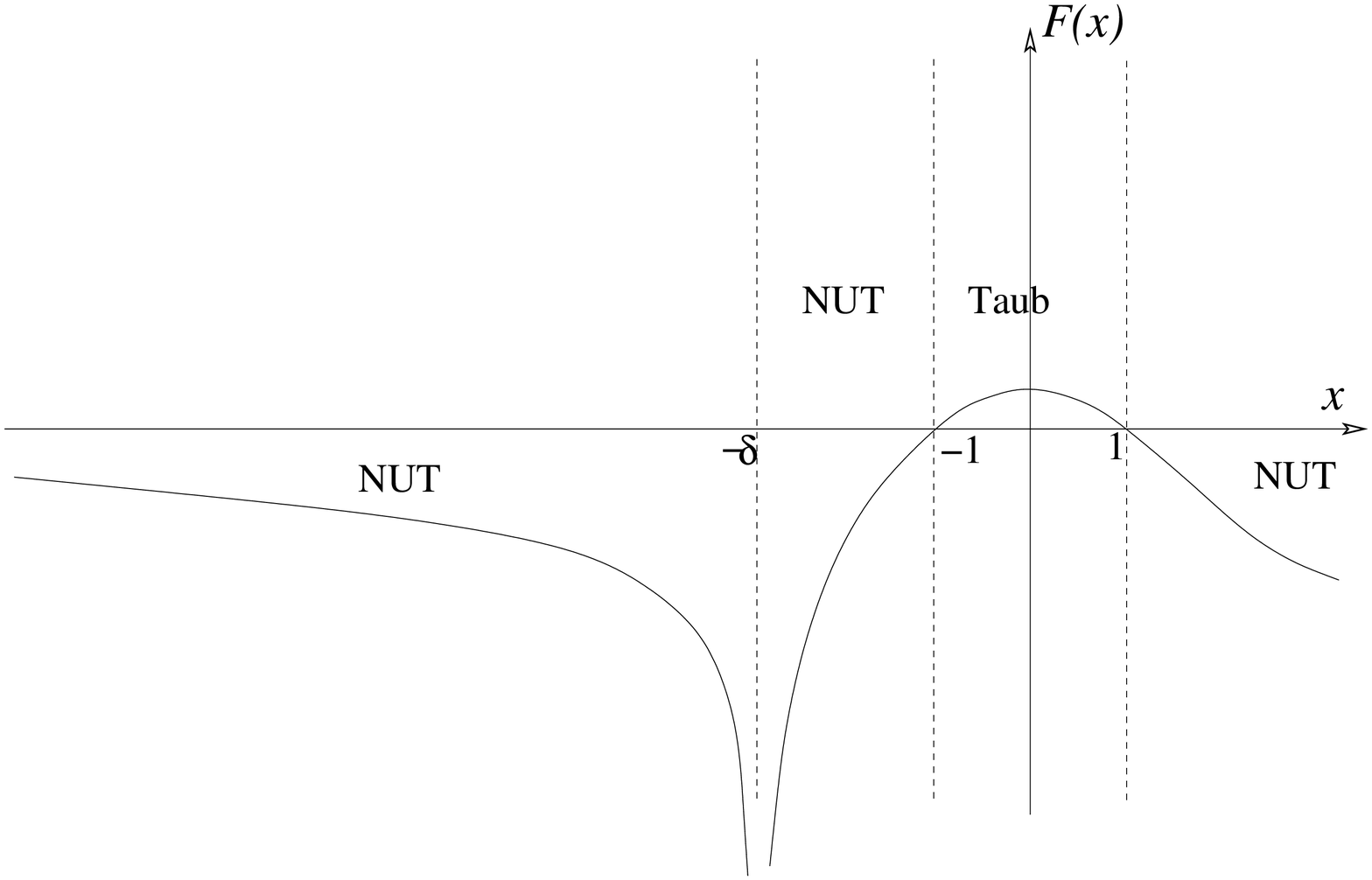}
\end{center}
\caption{\small The various regions in the stringy Taub--NUT geometry. There are two NUT regions, containing CTCs, and a Taub region. which is a cosmology. Note that there is a curvature singularity in the second NUT region, when $x=-\delta$.}
\label{zeros}
\end{figure}

The low--energy metric of the stringy Taub--NUT spacetime was
presented in ref.\cite{Johnson:1994jw} as (in string frame):
\begin{equation}
ds^2=k\left\{d\sigma^2 -\frac{\cosh^2\sigma -1}{(\cosh\sigma+\delta)^2}(dt-\lambda\cos\theta d\phi)^2+d\theta^2+\sin^2\theta d\phi^2\right\}\ ,
  \label{eq:taubnutlowenergy}
\end{equation}
where $0\leq\sigma\leq\infty$, $\delta\geq1$, $\lambda\geq0$. The
dilaton behaves as: 
\begin{equation}
  \Phi-\Phi_0=-\frac{1}{2}\ln(\cosh\sigma+\delta)\ ,
  \label{eq:dilaton}
\end{equation}
and there are other fields which we will discuss later.  This is in
fact the NUT region of the geometry, and $\sigma=0$ is a Misner
horizon.  We note here that the embedding presents a natural analytic
extension of this model which recovers the other NUT region and the
Taub cosmology as well: Replace $\cosh\sigma$ with the coordinate $x$:
\begin{equation}
ds^2=k\left( \frac{dx^2}{x^2-1} -\frac{x^2 -1}{(x+\delta)^2}(dt-\lambda\cos\theta d\phi)^2+d\theta^2+\sin^2\theta d\phi^2\right)\ ,
  \label{eq:taubnutlowenergyextended}
\end{equation}
with
\begin{equation}
\Phi-\Phi_0=-\frac{1}{2}\ln(x+\delta)\ ,
  \label{eq:dilatonextended}
\end{equation}
where now $-\infty\leq x\leq+\infty$. The three ranges of interest are
$1\leq x\leq+\infty$, ($x=\cosh\sigma$) which is the first NUT region
above, $-\infty\leq x\leq-1$ ($x=-\cosh\sigma$) which is a second NUT
region, and $-1\leq x\leq+1$ ($x=-\cos\tau$), which is a Taub region with
a Big Bang at $\tau=0$ and a Big Crunch at $\tau=\pi$. We shall see
shortly that this embedding is very natural from the point of view of
string theory, since $x$ is a natural coordinate on the group
$SL(2,\IR)$, which plays a crucial role in defining the complete theory.
It is interesting to sketch the behaviour of the function
$G_{tt}=F(x)=(1-x^2)/(x+\delta)^2$.  This is done in
figure~\ref{zeros}. Note that $F(x)$ vanishes at $x=\pm1$ and so for
$x=1-\tau$ where $\tau$ is small, the metric of the $(\tau,\xi)$ space
is:
\begin{equation}
ds^2=k\left(-(2\tau)^{-1}d\tau^2+\frac{2\tau}{(1+\delta)^2}d\xi^2\right)\ ,
  \label{eq:misnerlike1}
\end{equation}
which is of Misner form, and so the essential features of the
Taub--NUT spacetime persist in this stringy version of the spacetime.
Note that, unlike General Relativity's Taub--NUT solution, there is a
genuine curvature singularity in the metric, and it is located at
$x=-\delta$. The dilaton diverges there, and hence the string theory
is strongly coupled at this place, but it is arbitrarily far from the
regions of Misner space connecting the Taub and NUT regions, so we
will not need to worry about this locus for the questions of interest
in this paper.

Note that the $(x,t)$ plane is fibered over a family of $S^2$s which
have {\it constant} radius, as opposed to a radius varying with $x$.
This does not mean that we lose  key features of the geometry,
since {\it e.g.} in the Taub region, we still have a cosmology in
which the universe has $S^3$ topology, but its volume is controlled
entirely by the size of the circle fibre $(dt-\lambda\cos\theta
d\phi)$, which ensures that the universe's volume vanishes at the
beginning and the end of the cosmology.

\begin{figure}[ht]
\begin{center}
\includegraphics[scale=0.5]{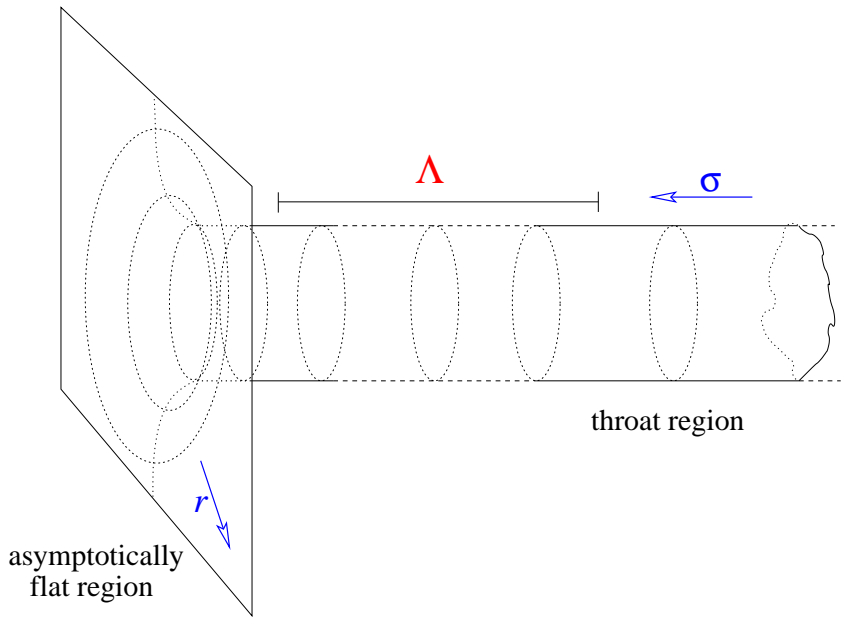}
\end{center}
\caption{\small A schematic showing the asymptotically flat region connected to the throat region located near the horizon at extremality. In the extremal limit, the typical measure, $\Lambda$ of the distance from a point on the outside to a point near the  horizon region   diverges logarithmically, and the throat region is infinitely long. The coordinate $\sigma$ is used  for the exact throat region in low--energy metric~\reef{eq:taubnutlowenergy}, while $r$ is the coordinate for the general low energy metric~\reef{eq:taubnutlow}.}
\label{throat}
\end{figure}

The constancy of the $S^2$s is in fact a feature, not a bug. It allows
the geometry to be captured in an exact conformal field theory, as we
shall recall in the next section. This geometry is the
``near--horizon'' limit of a  spacetime constructed as confirmation of
the statement in ref.\cite{Johnson:1994jw} that the metric in question
is indeed obtainable from the original Taub--NUT metric in a series of
steps using the symmetries of the heterotic string theory
action\cite{Johnson:1994ek,Kallosh:1994ba,Johnson:1994nj}. This
geometry is, in string frame:
\begin{equation}
ds^2=(a^2+f_2^2)\left\{
-\frac{f_1}{f_2^2}(dt+(\rho+1)l\cos\theta d \phi)^2
+f_1^{-1}dr^2+(r^2+l^2)(d\theta^2+\sin^2\theta d\phi^2)\right\}\ ,
  \label{eq:taubnutlow}
\end{equation}
where $f_1$ is as before, $\rho^2\geq1$ and 
\begin{equation}
f_2=1+(\rho-1)\frac{Mr+l^2}{r^2+l^2}\ ,\qquad{\rm and }\qquad a =(\rho-1)l\frac{r-M}{r^2+l^2}\ .
  \label{eq:functions}
\end{equation}
This metric has the full asymptotically flat part of the geometry and
connects smoothly onto the throat region, which develops in an
``extremal'' limit (analogous to that taken for charged black holes).
Figure~\ref{throat} shows a cartoon of this.  The
metric~\reef{eq:taubnutlowenergy} is obtained from it in the extremal
limit 
$\rho\to\infty, M\to0,l\to0$, 
where $m=\rho M$ and
$\ell=\rho l$ are held finite. The limit is taken in the neighbourhood
of $f_1=0$, and $\sigma$ is the scaled coordinate parameterizing $r$
in that region. 
The coordinate $t$ has to be rescaled as well to get matching expressions.
The parameters of metric~\reef{eq:taubnutlowenergy}
are recovered as: $\lambda=l/m$ and $\delta^2=1+l^2/M^2$. 

The stringy embedding giving rise to the metric~\reef{eq:taubnutlow}
(we have not displayed the other fields of the solution here) is
carried out starting from the metric~\reef{eq:taubnut} as follows:
(The details are in
refs.\cite{Johnson:1994ek,Kallosh:1994ba,Johnson:1994nj}). First, an
$O(1,1)$ boost (a subgroup of the large group of perturbative
non--compact symmetries possessed by the heterotic theory) is used to
generate a new solution, mixing the~$t$ direction with a $U(1)$ gauge
direction. This generates a gauge field $A_t$, a non--trivial dilaton,
and since there is a coupling of $t$ to $\phi$ in the original metric,
a gauge field $A_\phi$ and an anti--symmetric tensor background
$B_{t\phi}$. So the solution has electric and magnetic charges under a
$U(1)$ of the heterotic string, and non--trivial axion and dilaton
charge. We will not need the forms of the fields here. It turns out
that the dilaton has a behaviour which is ``electric'' in its
behaviour in a sense inherited from the behaviour of charged dilaton
black holes: It decreases as one approaches the horizon. Such holes do
not support the development of throats in the string frame metric, but
their ``magnetic'' cousins, where the dilaton has the opposite
behaviour, do support throats\footnote{In fact, an exact conformal
  field theory can be written for pure magnetic dilaton black holes in
  four dimensions\cite{Giddings:1993wn}, and it can be realized as a
  heterotic coset model as well\cite{Johnson:1994jw}.}. Using the
$SL(2,\IR)$ S--duality of the four dimensional effective action of the
heterotic string, which combines an electric--magnetic duality with an
inversion of the axi--dilaton field $\tau=a+ie^{-\Phi}$, a solution
with ``magnetic'' character can be
made\cite{Johnson:1994ek,Kallosh:1994ba}, which supports a throat in
the string frame metric. This is the solution whose metric we have
displayed in equation~\reef{eq:taubnutlow}.

So in summary, there is an embedding of General Relativity's
celebrated Taub--NUT solution into heterotic string theory which
preserves all of the interesting features: the NUT regions containing
CTCs, and the Taub region with its Big Bang and Big Crunch cosmology,
and (crucially) the Misner regions connecting them. There is a throat
part of the geometry which decouples from the asymptotically flat
region in an extremal limit, but which captures all of the features of
the Taub--NUT geometry of interest to us here.

The next thing we need to recall is that this throat geometry arises
as the low energy limit of a complete description in terms of a
conformal field theory, as presented in
ref.\cite{Johnson:1994jw}. 

\section{Exact Conformal Field Theory}
\subsection{The Definition}
\label{sec:exact1}
In ref.\cite{Johnson:1994jw}, the ``heterotic coset model'' technique
was presented, and one of the examples of the application of the
method was the model in question, from which the low energy metric in
equation~\reef{eq:taubnutlowenergyextended} was derived, for
$x=\cosh\sigma$. The other regions that have been presented here
(making up $-\infty\leq x\leq 0$) are easily obtained from the same
conformal field theory by choosing different coordinate patches in the
parent model, as we shall see.

Actions can be written for a large class of conformal field theories
obtained as coset
models\cite{Bardakci:1971nb,Halpern:1971ay,Goddard:1985vk,Goddard:1986ee,Kac:1985dv},
by using gauged WZNW
models\cite{Bardakci:1988ee,Gawedzki:1988hq,Gawedzki:1989nj,Karabali:1989au,Bowcock:1989xr,Karabali:1990dk}.
The ungauged model\cite{Witten:1984ar,Polyakov:1983tt} has some global
symmetry group $G$ which defines a conformal field
theory\cite{Knizhnik:1984nr,Gepner:1986wi,Goddard:1986bp} with an
underlying current algebra, and coupling it to gauge fields charged
under a subgroup $H\subset G$ gives the coset.  Such models have been
used to generate conformal field theories for many studies in string
theory, including cosmological contexts (see the introduction for some
references). It is important to note that the vast majority of these
models use a particular sort of gauging. The basic world--sheet field
is group valued, and we shall denote it as $g(z,{\bar z})$. The full
global invariance is $G_L\times G_R$, realized as: $g(z,\zb)\to
g_L^{\phantom{1}} g(z,\zb)g_R^{-1}$, for $g_L,g_R\in G$. The sorts of
group actions gauged in most studies are $g\to h_L^{\phantom{1}} g
h_R^{-1}$, for $h_L,h_R\in H$, and it is only a restricted set of
choices of the action of $h_{L}$ and $h_R$ which allow for the writing
of a gauge invariant action.  These are the ``anomaly--free''
subgroups, and the typical choice that is made is to correlate the
left and right actions so that the choice is essentially left--right
symmetric. This also gives a symmetric structure on the world sheet,
as appropriate to bosonic strings and to superstrings if one considers
supersymmetric WZNW models.  For these anomaly--free subgroups, a
gauge extension of the basic WZNW action can be written which is
$H$--invariant, and the resulting conformal field theory is
well--defined. The supersymmetric models can of course be turned into
heterotic string theories too, by simply tensoring with the remaining
conformal field theory structures needed to make a left--right
asymmetric model.

The general heterotic coset model goes beyond this, and exploits the
basic fact that the heterotic string is asymmetric in how it is built.
The idea is to allow oneself the freedom to choose to gauge far more
general subgroups. This might well produce anomalies, but permits one
to choose to retain certain global symmetries which might be of
interest (such as spacetime rotations) and/or use in the conformal
field theory.  Introducing right--moving fermions to achieve a
right--moving supersymmetry is easy to do, and they contribute extra
terms to the anomaly, making matters worse in general: Their couplings
(the effective charges they carry under $H$) are completely determined
by supersymmetry, so one has no choice.  Of course, one does not have
a well--defined model if there are anomalies, so ultimately they must
be eliminated. This is achieved as follows\cite{Johnson:1994jw}. Note
that the left--moving fermions can be introduced with {\it arbitrary}
couplings (charges under $H$), since there is no requirement of
left--moving supersymmetry in the heterotic string. The anomaly they
contribute comes with the opposite sign to that of the others, since
they have the opposite chirality.  The requirement that the anomaly
cancels can be satisfied, since it just gives a set of algebraic
equations to solve for the charges. The resulting model is a conformal
field theory with (0,1) world--sheet supersymmetry, (enhanced to
$(0,2)$ when $G/H$ is
K\"ahler\cite{Banks:1988cy,Sen:1986mg,Witten:1992mk}) naturally
adapted to the heterotic string.

It is important to note that the types of heterotic models obtained by
this method are very different from the types of models obtained by
gaugings that do not cancel the anomalies against those of the gauge
fermions. One way to see the difference is to note that since the
anomaly is proportional to $k$, the cancellation equation puts the
gauge charge at the same order as the metric. This means that there is
a non--trivial modification of the geometry one would read off from
the WZNW action, traceable to the left--moving fermions. We will
explain this more shortly.

By way of example, we simply present the model relevant to our study
here\cite{Johnson:1994jw}. The group in question is $SL(2,\IR)\times
SU(2)$, and the group elements are denoted $g_1$ and $g_2$
respectively. Let the levels of the models be denoted $k_1$ and $k_2$,
respectively.  We are interested in a $U(1)_A\times U(1)_B$ subgroup
($A$ and $B$ are just means of distinguishing them) which acts as
follows:
\begin{eqnarray}
U(1)_A\times U(1)_B:\left\{\begin{matrix}g_1&\to& e^{\epsilon_A\sigma_3/2}g_1e^{(\delta\epsilon_A+\lambda\epsilon_B)\sigma_3/2}\cr g_2&\to& g_2 e^{i\epsilon_B\sigma_3/2}\end{matrix}\right.\,
  \label{eq:groupaction}
\end{eqnarray}
Notice that there is a whole global $SU(2)_L$ of the original
$SU(2)_L\times SU(2)_R$ untouched. This is a deliberate choice to give
a model with spacetime $SU(2)$ invariance (rotations) in the end. With
that, and the other asymmetry introduced by the presence of $\lambda$
and $\delta$, the gauging is very anomalous. Once right--moving
supersymmetry fermions are introduced, the anomalies are proportional
to $-k_1(1-\delta^2)+2\delta^2$ from the $AA$ sector,
$k_1\delta\lambda+2\delta\lambda$ from the $AB$ sector, and
$k_2+k_1\lambda^2+2(1+\lambda^2)$ from the $BB$ sector. The
$k$--independent parts come from the fermions. Next, four left--moving
fermions are introduced. Two are given charges $Q_{A,B}$ under
$U(1)_{A,B}$ and the other two are given charges $P_{A,B}$. Their
anomalies are $-2(Q_A^2+P_A^2)$, $-2(Q_AQ_B+P_AP_B)$, and
$-2(Q_B^2+P_B^2)$, respectively, from the various sectors $AA$, $AB$,
$BB$. So we can achieve an anomaly--free model by asking that:
\begin{eqnarray}
-k_1(1-\delta^2)&=& 2(Q_A^2+P_A^2-\delta^2)\nonumber\\
k_1\delta\lambda&=& 2(Q_AQ_B+P_AP_B-\delta\lambda)\nonumber\\
k_2+k_1\lambda^2&=& 2(Q_B^2+P_B^2-(1+\lambda^2)) \ .
  \label{eq:anomalies}
\end{eqnarray}
It is a highly non--trivial check on the consistency of the model to
note that in the solution--generating techniques used to verify the
observation made in ref.\cite{Johnson:1994jw} that our stringy
solution~\reef{eq:taubnutlowenergy} can be obtained from the basic
Taub--NUT solution~\reef{eq:taubnut}, the charges in the resulting
throat metric turn out to be given in terms of the parameters $M,l$
and $\rho$ in such a way that they satisfy the anomaly equations
above, in the large $k$ limit (which is appropriate to low--energy).
See ref.\cite{Johnson:1994ek}.

The central charge of this four dimensional model is:
\begin{equation}
c=\frac{3k_1}{k_1-2}+\frac{3k_2}{k_2+2}\ ,
  \label{eq:centralcharge}
\end{equation}
where the $-2$ from gauging is cancelled by the $+2$ from four bosons
on the left and right.  We can ask that this be equal to $6$, as is
appropriate for a four dimensional model, tensoring with another
conformal field theory to make up the internal sector, as
desired\footnote{Actually, we can also choose other values of $c$, and
  adjust the internal theory appropriately.}.  The result is that
$k_1=k_2+4$.

In ref.\cite{Johnson:1994jw}, the metric for the throat region was
discovered by working in the low energy limit where $k_1$ and $k_2$
are large, and denoted simply as $k$. In this paper, we study the case
of going beyond this large $k$ (low energy) approximation and derive
the geometry which is correct to all orders in the $\alpha^\prime\sim
1/k$ expansion.

\subsection{Writing The Full Action}
\label{sec:exact2}
The $G=SL(2,\IR)\times SU(2)$ WZNW model is given by:
\begin{equation}
S(g_1,g_2)=-k_1I(g_1)+k_2I(g_2)\ ,
\label{eq:bigwznw}
\end{equation}
where 
\begin{equation}
I(g)=-\frac{1}{4\pi}\int_\Sigma d^2z \Tr(g^{-1}\partial_z g g^{-1}\partial_{\bar z}g)-i\Gamma(g)\ ,
  \label{eq:wznw}
\end{equation}
with
\begin{equation}
\Gamma(g)=\frac{1}{12\pi}\int_{\cal B}d^3\sigma \epsilon^{abc}\Tr(g^{-1}\partial_a g g^{-1}\partial_b g g^{-1}\partial_c g)\ .
  \label{eq:wzterm}
\end{equation}
The group valued fields $g_1(z,{\bar z})\in SL(2,\IR)$ and
$g_2(z,{\bar z})\in SU(2)$ map the world--sheet $\Sigma$ with
coordinates $(z,{\bar z})$ into the group $SL(2,\IR)\times SU(2)$.
Part of the model is defined by reference to an auxiliary spacetime
${\cal B}$, whose boundary is $\Sigma$, with coordinates $\sigma^a$.
The action $\Gamma(g)$ is simply the pull--back of the $G_L\times G_R$
invariant three--form on $G$.

With reference to the $U(1)_A\times U(1)_B$ action chosen in
equation~\reef{eq:groupaction}, the gauge fields are introduced with
the action:
\begin{eqnarray}
S(g_1,g_2,A)&=&\frac{k_1}{8\pi}\int d^2z\Biggl\{-2(\delta A_\zb^A+\lambda A^B_\zb)\Tr[\sigma_3g_1^{-1}\partial_z g_1]-2A_z^A\Tr[\sigma_3\partial_\zb g_1g_1^{-1}]\nonumber\\
&&\hskip2cm+A_z^AA_\zb^A(1+\delta^2+\delta\Tr[\sigma_3g_1\sigma_3g_1^{-1}])+\lambda^2A_z^BA_\zb^B\nonumber\\
&&\hskip2cm+\lambda\delta A_z^A A_\zb^B+A_z^B A_\zb^A(\lambda\delta+\lambda\Tr[\sigma_3g_1\sigma_3g_1^{-1}])
 \Biggr\}\nonumber\\
&&\hskip1cm+\frac{k_2}{8\pi}\int d^2 a\left\{2iA_\zb^B\Tr[\sigma_3g_2^{-1}\partial_z g_2]+A_z^BA_\zb^B\right\}\ ,
  \label{eq:gaugeextension}
\end{eqnarray}
and we note that we have written the generators as
\begin{equation}
t^{(1)}_{A,R}=-\delta\frac{\sigma_3}{2},\quad t^{(1)}_{A,L}=\frac{\sigma_3}{2},\quad
t^{(1)}_{B,R}=-\lambda\frac{\sigma_3}{2},\quad t^{(2)}_{B,R}=-i\frac{\sigma_3}{2}\ .
  \label{eq:generators}
\end{equation}
The anomaly under variation $\delta
A^{A(B)}_a=\partial_a\epsilon_{A(B)}$ can be written as:
\begin{equation}
{\cal A}_{ab}=\frac{1}{4\pi}\Tr[t_{a,L}t_{b,L}-t_{a,R}t_{b,R}]\epsilon_a\int\! d^2z F^b_{z\zb}\ ,
  \label{eq:anomaly}
\end{equation}
(no sum on $a,b$) and we've defined  $\Tr=-k_1\Tr_1+k_2\Tr_2$. The right--moving
fermions have an action:
\begin{equation}
I_R^F=\frac{i}{4\pi}\int\! d^2 z\Tr(\Psi_R{\cal D}_z\Psi_R)\ ,
  \label{eq:rightfermions}
\end{equation}
where $\Psi_R$ takes values in the orthogonal complement of the Lie
algebra of $U(1)_A\times U(1)_B$, (so there are four right--movers, in
fact) and
\begin{equation}
{\cal D}_z\Psi_R=\partial_z\Psi_R-\sum_a A^a_\zb[t_{a,R},\Psi_R]\ ,
  \label{eq:covariant}
\end{equation}
The four left--moving fermions have action:
\begin{equation}
I_L^F=-\frac{ik_1}{4\pi}\int\! d^2z\left\{\lambda^1_L[\partial_z+Q_AA_z^A+Q_BA_z^B]\lambda^2_L\right\}+\frac{ik_2}{4\pi}\int d^2 z\left\{\lambda_L^3[\partial_z+P_AA_z^A+P_B A_z^B]\lambda_L^4\right\}\ .
  \label{eq:leftfermions}
\end{equation}
Under the gauge transformation $\delta
A^{A(B)}_a=\partial_a\epsilon_{A(B)}$, these two sets of fermion
actions yield the anomalies discussed earlier, but at one--loop, while
the WZNW model displays its anomalies classically. It is therefore
hard to work with the model in computing a number of properties. In
particular, in working out the effective spacetime fields it is useful
to integrate out the gauge fields. It is hard to take into account the
effects of the successful anomaly cancellation if part of them are
quantum and part classical. The way around this awkward state of
affairs\cite{Johnson:1994jw} is to bosonize the fermions. The
anomalies of the fermions then appear as classical anomalies of the
action. The bosonized action is:
\begin{eqnarray}
I_B&=&\frac{1}{4\pi}\int d^2 z\Biggl\{|\partial_z\Phi_2-P_AA_z^A-(P_B+1)A_z^B|^2+|\partial_z\Phi_1-(Q_B+\lambda)A_z^B-(Q_A+\delta)A_z^A|^2 
\nonumber\\
&&\hskip2cm-\Phi_1[(Q_B-\lambda)F_{z\zb}^B+(Q_A-\delta)F_{z\zb}^A]-\Phi_2[(P_B-1)F_{z\zb}^B+P_AF^A_{z\zb}]\nonumber\\
&&\hskip2cm+[A_\zb^A A_z^B-A_z^AA_\zb^B][\delta Q_B-\lambda Q_A -P_A]
   \Biggr\}\ ,
  \label{eq:bosonized}
\end{eqnarray}
which under variations:
\begin{equation}
\delta
A^{A(B)}_a=\partial_a\epsilon_{A(B)}\ ,\quad \delta\Phi_1=(Q_A+\delta)\epsilon_A+(Q_B+\lambda)\epsilon_B\ ,\quad
\delta\Phi_2=P_A\epsilon_A+(P_B+1)\epsilon_B\ ,
  \label{eq:variation}
\end{equation}
manifestly reproduces the anomalies presented earlier.

\subsection{Extracting the Low Energy Metric}

At this stage, it is possible to proceed to derive the background
fields at leading order by starting with the Lagrangian definition
given in the previous section and integrating out the gauge fields,
exploiting the fact that they appear quadratically in the action. As
these fields are fully quantum fields, this procedure is only going to
produce a result which is correct at leading order in the $1/k$
expansion, where $k$ is large. This is because we are using their
equations of motion to replace them in the action, and neglecting
their quantum fluctuations. Before turning to how to go beyond that,
let us note that there is an important subtlety even in the derivation
of the leading order metric. This is not an issue for coset models
that are not built in this particularly heterotic manner, and so is a
novelty that cannot be ignored.

The coordinates we use for $SL(2,\IR)$ and $SU(2)$ are:
\begin{equation}
g_1=\frac{1}{\sqrt{2}}\begin{pmatrix}
e^{t_+/2}(x+1)^{1/2}&e^{t_-/2}(x-1)^{1/2}\cr e^{-t_-/2}(x-1)^{1/2}&e^{-t_+/2}(x+1)^{1/2}\end{pmatrix}\ ,
  \label{eq:parameters1}
\end{equation}
where $t_\pm=t_L\pm t_R$, and $-\infty\leq t_R, t_L, x\leq\infty$, and
the Euler angles
\begin{equation}
g_2=\begin{pmatrix}
\phantom{-}e^{i\phi_+/2}\cos\frac{\theta}{2}&e^{i\phi_-/2}\sin\frac{\theta}{2}\cr -e^{-i\phi_-/2}\sin\frac{\theta}{2}&e^{-i\phi_+/2}\cos\frac{\theta}{2}\end{pmatrix}\ ,
  \label{eq:parameters2}
\end{equation}
where $\phi_\pm=\phi\pm\psi$, $0\leq\theta\leq\pi$,
$0\leq\psi\leq4\pi$, and $0\leq\phi\leq2\pi$. Note that the full range
of $x$ is available here, while remaining in $SL(2,\IR)$. In
ref.\cite{Johnson:1994jw}, the range $x=\cosh\sigma\geq 1$ was used.
The larger range reveals the connection to the Taub and the other NUT
region. This extension is very naturally inherited from the
$SL(2,\IR)$ embedding\footnote{See ref.\cite{Elitzur:2002rt} for a
  discussion of how an $SL(2,\IR)$ structure also provides a natural
  extension for the discussion of wavefunctions in related
  spacetimes.}.

The gauge we fix to before integrating out the gauge fields is:
\begin{equation}
t_L=0\ ,\quad \psi=\pm\phi\ ,
  \label{eq:gaugefix}
\end{equation}
where the sign choice depends on which coordinate patch we 
investigate, such that $+$ refers to the North pole on the $S^2$ 
parameterized by $(\theta,\phi)$ and $-$ refers to the South pole,
and we write $t_R=t$. One can then
read off various spacetime fields from the resulting $\sigma$--model,
by examining terms of the form $C_{ij}\partial_z \chi^i\partial_\zb
\chi^j$, where here $\chi^j$, is a place holder for any worldsheet
field, and $j$ denoted which field is present. When $i,j$ are such
that $\chi^i\chi^j$ run over the set of fields $t,x,\theta,\phi$, then
the symmetric parts of $C_{ij}$ give a metric we shall call
$G^0_{\mu\nu}$, and the antisymmetric parts give the antisymmetric
tensor potential $B_{\mu\nu}$. When $i,j$ are such that $\chi^i$ is
one of the bosonized fermions and $\chi^j$ is one of
$t,x,\theta,\phi$, the $C_{ij}$ is a spacetime gauge potential, either
from the (1) or the (2) sector: $A^{(1,2)}_\mu$.

Note that $G^0_{\mu\nu}$ is {\it not} the correct spacetime metric at
this order. This is a crucial point\cite{Johnson:1994jw}. The anomaly
cancellation requirement means that the contribution from the
left--movers has a significant modification to the naive metric. The
most efficient way of seeing how it is modified is to re--fermionize
the bosons, using as many symmetries as one can to help in deducing
the normalization of the precise couplings. After some
work\cite{Johnson:1994jw}, it transpires that the correct metric (to
leading order) is:
\begin{equation}
G_{\mu\nu}=G^0_{\mu\nu}-\frac{1}{2k}[A^1_\mu A^1_\nu+A^2_\mu A^2_\nu]\ ,
  \label{eq:correctmetric}
\end{equation}
where it can be seen that because $A\sim Q$ and from the anomaly
equations~\reef{eq:anomalies} we have $Q\sim \sqrt{k}$, this gives a
non--trivial correction to the metric one reads off naively. This is
the clearest sign that these heterotic coset models are quite
different from coset models that have commonly been used to make
heterotic string backgrounds by tensoring together ordinary cosets. In
those cases, typically $A\sim Q\sim 1$ and so at large $k$, the
correction is negligible.

This sets the scene for what we will have to do when we have
constructed the exact effective $\sigma$--model. We will again need to
correct the naive metric in a way which generalizes
equation~\reef{eq:correctmetric}, in order to get the right spacetime
metric.

\subsection{The Exact Effective Action}
\label{sec:effective1}
In the previous section, we treated the gauge fields as classical
fields, substituting their on--shell behaviour into the action to
derive the effective $\sigma$--model action for the rest of the fields
and ignoring the effects of quantum fluctuations arising at subleading
order in the large $k$ expansion. To include all of the physics and
derive a result valid at any order in $k$, we need to do better than
this. For ordinary coset models, this sort of thing has been achieved
before, using a number of methods. To our knowledge, this was first
done in ref.\cite{Dijkgraaf:1992ba} in the context of the
$SL(2,\IR)/U(1)$ coset model studied as a model of a two--dimensional
black hole\cite{Witten:1991yr}. The exact metric and dilaton were
written down by appealing to a group theoretic argument, writing the
exact expressions for the quadratic Casimirs for $G$ and for $H$, in
terms of the target space ($G/H$) fields, and then equating their
difference to the Laplacian for the propagation of a massless field
(the tachyon) in the background. The proposed metric and dilaton were
verified at higher orders by explicit calculation in
ref.\cite{Tseytlin:1991ht,Jack:1993mk}, and the argument was
generalized and applied to a number of other models in a series of
papers\cite{Bars:1992sr,Bars:1993dx}. An elegant alternative method
was developed in refs.\cite{Tseytlin:1993ri,Bars:1993zf}, and is the
one we adapt for use here. We must extend it to work for the heterotic
coset models, since although heterotic backgrounds are considered in
some of those works, they are of the mildly heterotic type which are
essentially similar to the superstring models: an asymmetric
arrangement of fermions is merely tensored in as dressing.

Since there will be a fair amount of messy computation in what
follows, we state the key ideas in what follows: It is
known\cite{Leutwyler:1992tv,Tseytlin:1993ri,Shifman:1991pa} that the
exact effective action for the WZNW model defined in ref.\reef{eq:wznw}
is extremely simple to write down.  One takes the form of the basic
action at level $k$, $kI(g)$, where $g$ is a quantum field, and one
writes for the full quantum effective action $(k-c_G)I(g)$, where now
$g$ should be taken as a classical field, and $c_G$ is the dual
Coxeter number of the group~$G$. This is particularly simple since $k$
only enters the action as an overall multiplicative factor, which then
gets shifted. The key observation of
refs.\cite{Tseytlin:1993ri,Bars:1993zf} is that this can be applied to
a gauged WZNW model as well, by exploiting the fact that if one writes
$A_z=\partial_z h_zh^{-1}_z$ and $A_\zb=\partial_\zb h_\zb
h^{-1}_\zb$, the action can be written as the sum of two formally
decoupled WZNW models, one for the field $g^\prime=h_\zb^{-1}gh_z$ at
level $k$ and the other for the field $h^\prime=h_\zb^{-1}h_z$ at
level $2c_H-k$.  To write the exact effective action, one shifts the
levels in each action: $k\to k-c_G$ and $2c_H-k\to 2c_H-k-c_H=c_H-k$,
and treats the fields as classical.  Transforming back to the original
variables, one gets the original gauged WZNW model with its level
shifted according to $k\to k -c_G$, together with a set of new terms
for $A_z,A_\zb$ which are proportional to $c_H-c_G$, and have no $k$
dependence.  Because there is no multiplicative factor of $k$ in these
new terms, it is easy to see that the large $k$ contribution to the
result of integrating out the gauge fields will be the same as before.
For results exact in $k$, there will be a family of new contributions
to the $\sigma$--model couplings upon integrating out the gauge
fields. In this effective action, they are to be treated as classical
fields now and so once the integration is done, there are no further
contributions from quantum fluctuations to take into account.  The
metrics derived using this method are the same as those constructed
using the algebraic approach, which is a useful consistency
check\cite{Tseytlin:1993ri,Bars:1993zf}.

Note that the new pieces in the effective action are non--local in the
fields $A_z, A_\zb$ (although local in the $h_z,h_\zb$). This
difficulty does not present a problem for the purposes of reading off
the spacetime fields, since it is enough to work in the zero--mode
sector of the string to capture this information. This amounts to
dropping all derivatives with respect to $\sigma$ on the world--sheet
and working with the reduced ``point--particle'' Lagrangian for that
aspect of the computation\cite{Bars:1993zf}.

Let us turn to the model in question. Here, we exploit the
fact\cite{Johnson:1994jw,Johnson:1995kv,Johnson:1994kh} that our
heterotic coset model, in its bosonized form (where all the anomalies
are classical) can be thought of as an asymmetrically gauged WZNW
model for $G/H$ supplemented by another asymmetrically gauged WZNW
model for $SO({\rm dim}\,\, G-{\rm dim }\,\,H)/H$, representing the
fermions.  We should be able to carry out a similar set of changes of
variables to write the whole model as a set of decoupled WZNW models,
transform to the effective action, and then rewrite it back in the
original variables to see what new terms the effective action supplies
us with.  Then we have to integrate out the gauge fields and
---crucially--- correctly re--fermionize the bosons to read off the
spacetime fields. This is the subject of the next subsection.
The reader wishing to skip to the result can pick up the story again
at the beginning of  subsection~\ref{sec:properties}.

\subsection{Computation of the Exact Effective Action}
\label{sec:effective2}

As noted above, the fermions can also be represented as a gauged WZNW
model based on the coset $SO(D)/H$, with $D={\rm dim}~G-{\rm dim}~H=6-2=4$.
Doing this, the complete classical action can
be written as:
\begin{equation}
  S = -k_1 I(g_1) + k_2 I(g_2) + I(g_f)\ ,
\end{equation}
with $g_1\in SL(2,\mathbb{R})$, $g_2\in SU(2)$, and $g_f\in SO(4)$.
It is convenient to write
\begin{equation}
  g = \left(\begin{array}{ccc} 
      g_1 & 0 & 0
      \\ 0 & g_2 & 0
      \\ 0 & 0 & g_f
  \end{array}\right) \in SL(2,\mathbb{R})\times SU(2)\times SO(4)\ .
\end{equation} 
To gauge the subgroup $H=U(1)_A\times U(1)_B$ we introduce the
covariant derivative
\begin{equation}
  \mathcal{D}_\mu g = \partial_\mu g 
  +A^a_{\mu,L}~g - g~A^a_{\mu,R},
\end{equation}
where $A_{\mu,L} = A^a_\mu t_{a,L}$ and 
$A_{\mu,R} = A^a_\mu t_{a,R}$. 
These are the gauge fields, which take values in the Lie
algebra of $H$. 
With $f_L\in H_L$, $f_R\in H_R$, the gauge transformation is written
\begin{equation}
  g \to f_L g f_R^{-1}\ .
\end{equation}
The $t_{a,L}$ are left generators, and
$t_{a,R}$ are right generators of $H$. 
Using the block diagonal notation above, we can write
\begin{equation}
  A = A^a \left(\begin{array}{ccc} 
      t^{(1)}_a & 0  & 0
      \\ 0 & t^{(2)}_a & 0
      \\ 0 & 0 & t^{(f)}_a
  \end{array} \right) \in {\rm Lie}(H)\ ,
\end{equation}
where $t^{(1)}_a$ and $t^{(2)}_a$ are $2\times 2$ matrices,
and $t^{(f)}_a$ are $4\times 4$ matrices.

The gauged WZNW model is
\begin{equation}
  S_{gWZNW} = -k_1\bigl[ I(g_1) + S_1(g_1,A) \bigr] 
            +k_2\bigl[ I(g_2) + S_1(g_2,A) \bigr]
            +\bigl[ I(g_f) + S_1(g_f,A) \bigr]\ ,
\end{equation}
where
\begin{align}
  S_1(g,A) &= \frac{2}{4\pi}\int d^2\!z \Tr \Bigl\{
    A_{\bz,L} \partial_z g g^{-1} - A_{z,R} g^{-1}\partial_{\bz} g
    - A_{\bz,L} g A_{z,R} g^{-1} + \half(A_{z,L} A_{\bz,L} +
    A_{z,R}A_{\bz,R}) \Bigr\}\ .
\end{align}
Since there is no gauge-invariant extension for the Wess--Zumino term
$\Gamma(g)$ for general subgroup~$H$, this action has (in general)
classical anomalies. However, there is a unique extension such that
the anomalies do not depend on $g$, but only on gauge
fields\cite{Witten:1992mm}. This extension has been used in the
expression above.

\subsubsection{A Change of variables}
By the change of variables
\begin{equation}
\begin{array}{ll}
  A_{z,L} = -\partial_z h_z h_z^{-1}\ , & 
    A_{z,R} = -\partial_z \tilde{h}_z \tilde{h}_z^{-1}\ ,
    \qquad h, \tilde{h} \in H\ ,
    \\
  A_{\bz,L} = -\partial_\bz h_\bz h_\bz^{-1}\ , &
    A_{\bz,R} = -\partial_\bz \tilde{h}_\bz \tilde{h}_\bz^{-1}\ ,
\end{array}
\end{equation}
we find
\begin{equation}
\begin{split}
  S_1(g,h) = \frac{2}{4\pi} \int d^2\!z \Tr\Bigl\{&
    -\partial_z g g^{-1} \partial_\bz h_\bz h^{-1}_\bz
    + g^{-1}\partial_\bz g \partial_z \tilde{h}_z \tilde{h}^{-1}_z
    - \partial_\bz h_\bz h^{-1}_\bz g \partial_z \tilde{h}_z
  \tilde{h}_z^{-1} g^{-1}
  \\ &
  +\half( \partial_\bz h_\bz h_\bz^{-1} \partial_z h_z h_z^{-1}
         +\partial_\bz \tilde{h}_\bz \tilde{h}_\bz^{-1} 
            \partial_z \tilde{h}_z \tilde{h}_z^{-1} )
  \Bigr\}\ .
\end{split}
\end{equation}
The Polyakov-Wiegmann identity \cite{Polyakov:1983tt} leads to the identities:
\begin{align}
  \begin{split}
  I(h^{-1}_\bz g \tilde{h}_z) =& I(g)+I(h_\bz^{-1})+I(\tilde{h}_z)
  \nonumber \\&
    +\frac{2}{4\pi}\int d^2\!z \Tr\Bigl[
    -\partial_\bz h_\bz h_\bz^{-1} \partial_z g g^{-1}
    -\partial_\bz h_\bz h_\bz^{-1} g\partial_z\tilde{h}_z\tilde{h}_z^{-1}g^{-1}
    +g^{-1}\partial_\bz g \partial_z\tilde{h}_z\tilde{h}_z^{-1}
    \Bigr]\ ,
  \end{split}
   \nonumber  \\
  I(h_\bz^{-1}h_z) = & I(h_\bz^{-1})+ I(h_z)
    +\frac{2}{4\pi}\int d^2\!z \Tr\Bigl[
    -\partial_\bz h_\bz h_\bz^{-1} \partial_z h_z h_z^{-1}
    \Bigr]\ ,
   \nonumber  \\
  I(\tilde{h}_\bz^{-1}\tilde{h}_z)  =& I(\tilde{h}_\bz^{-1})+I(\tilde{h}_z)
    +\frac{2}{4\pi}\int d^2\!z \Tr\Bigl[
    -\partial_\bz\tilde{h}_\bz\tilde{h}_\bz^{-1} 
     \partial_z \tilde{h}_z \tilde{h}_z^{-1}
    \Bigr]\ .
\end{align}
Using these, the classical action can be written as\footnote{In this
  case of Abelian $H$, the Jacobian for the change of variables
  vanishes.}:
\begin{equation}
\begin{split}
  S_1  &=-I(g) + I(h_\bz^{-1}g\tilde{h}_z)
  -\half \Bigl[I(h_\bz^{-1}h_z)+I(\tilde{h}_\bz^{-1}\tilde{h}_z)\Bigr]
  -\half C
  \Bigr]\ ,\nonumber\\
\mbox{\rm where}\quad C&\equiv I(h_\bz^{-1})-I(\tilde{h}_\bz^{-1})-I(h_z)+I(\tilde{h}_z)\ .
\end{split}
\end{equation}
The term $C$ is not manifestly gauge invariant, but the others are.
Note that if $A_L=A_R$, then $C=0$, in which case the gauging is
classically anomaly-free.  Otherwise, the anomalous terms $C_i$ may
look disturbing, but in fact they cancel, $\sum k_{(i)}C_i = 0$, as
will follow from the anomaly cancellation equations~\reef{eq:anomalies}.

Taking all this into account, we can write the  action as:
\begin{equation}
  S
= -\sum_{i=1,2,f} \Bigl\{
    k_{(i)} I({h_\bz^{-1}g_i\tilde{h}_z})
    -(k_{(i)}-2c_H)\half
      \Bigl[I({h_\bz^{-1}h_z}) 
      +I({\tilde{h}_\bz^{-1}\tilde{h}_z})\Bigr] 
    \Bigr\}\ ,
\end{equation}
with $k_{(1)}=k_1$, $k_{(2)}=-k_2$ and $k_{(f)}=-1$ and we note that
$h^{-1}_\zb g h_z\in G$, $h_\zb^{-1}h_z\in H$, and ${\tilde
  h}_\zb^{-1}{\tilde h}_z\in H$.  Now, as promised in the previous
section, we have achieved the rewriting of the full action in the form
of a sum of WZNW actions, which allows us to write
down the quantum effective action in a very simple way.

\subsubsection{Effective action}
Using the simple prescription given above,
\begin{equation}
\begin{split}
  &\text{for $G$: } k_{(i)} \to k_{(i)}-c_{G_i}\ , \qquad
  \\
  &\text{while for $H$: } -k_{(i)}+2c_H \to (-k_{(i)}+2c_H)-c_H = -(k_{(i)}-c_H)\ ,
\end{split}
\end{equation}
we find the effective action
\begin{equation}
\begin{split}
  S^{eff} = -\sum_{i=1,2,f} \Bigl\{
    (k_{(i)}-c_{G_i}) I(h_\bz^{-1}g_i\tilde{h}_z)
    -(k_{(i)}-c_H)\half
      \Bigl[I(h_\bz^{-1}h_z)+I(\tilde{h}_\bz^{-1}\tilde{h}_z)\Bigr] 
    \Bigr\}\ , 
\end{split}
\end{equation}
where $G_1=SL(2,\mathbb{R})$, $G_2=SU(2)$, $G_f=SO(4)$,
$H=U(1)\!\times\!U(1)$.  Again, the action is manifestly gauge
invariant.  It is important to note here that the level constant for
the fermionic
sector $k_{(f)}=1$ is \emph{not} shifted.

\subsubsection{Return to the  original variables} 
We now change variables back to the original ones, using the
identities given above.  We find
\begin{equation}
\begin{split}
  S^{\rm eff} = &-\sum_{i=1,2,f} \Bigl\{
    (k_{(i)}-c_{G_i}) \Bigl[
      I(g) + S_1(g,A) +\half\bigl[ I_2(A_L)+I_2(A_R)\bigr]
      +\half C_i
    \Bigr]\\
    &\hskip3cm -(k_{(i)}-c_H)\half
      \Bigl[I_2(A_L)+I_2(A_R)\Bigr] 
    \Bigr\}\ ,
\end{split} 
\end{equation}
where $I_2(A_L) \equiv I(h_\bz^{-1}h_z)$, $I_2(A_R)\equiv
I(\tilde{h}_\bz^{-1}\tilde{h}_z)$.  Observe that the $C_i$'s have come
back into the action.  Rewritten, this is
\begin{equation}
  S^{\rm eff} = -\sum_{i=1,2,f}
    (k_{(i)}-c_{G_i}) \Bigl[
      I(g) + S_1(g,A) -\frac{\lambda_i}{2}\bigl[ I_2(A_L)+I_2(A_R)\bigr]
      \Bigr]\ ,
\end{equation}
where $\lambda_i=\frac{c_{G_i}-c_H}{k_{(i)}-c_{G_i}}$.

\subsection{Extracting the Exact Geometry}
\label{sec:exactgeometry}
As we stated earlier, a problem with working with this action is that
it has terms which are non--local in the gauge fields. Since we are
going to integrate these out, this is inconvenient.  To avoid this
complication, we shall reduce to the zero mode
sector\cite{Bars:1993zf}, which is enough to extract the information
we want. The zero mode sector is obtained by letting fields depend on
worldsheet time only. So $\partial_z$ and $\partial_\bz \to
\partial_\tau$.  We also denote $A$ by $a$ in this limit.  This leads
to the desired simplifications.  Note the additional simplification
that the WZ part of the WZNW action vanishes in this sector, {\it
  i.e.,}  $\Gamma(g)\to 0$. 

The resulting action is
\begin{equation}
\begin{split}
  S^{\rm eff}_0 = -\sum \frac{(k_{(i)}-c_{G_i})}{4\pi}&\int d\tau \Bigl\{
    \Tr(g^{-1}\partial g g^{-1}\partial g)
    \\ &
    +2\Tr\bigl[a_{\bz,L}\partial g g^{-1} -a_{z,R} g^{-1}\partial g 
    -a_{\bz,L} g a_{z,R} g^{-1} + \half(a_{z,L} a_{\bz,L} +
    a_{z,R}a_{\bz,R}) \bigr]
    \\ &
    -\lambda_i \half \Tr \bigl[ (a_{\bz,L}-a_{z,L})^2 +
    (a_{\bz,R}-a_{z,R})^2 \bigr]
    \\ &
    +\half \Tr\bigl[ a_{z,R}a_{z,R} -
    a_{\bz,R}a_{\bz,R} + a_{\bz,L} a_{\bz,L} - a_{z,L} a_{z,L} \bigr]
  \Bigr\}\ .
\end{split}
\end{equation}
This is a local action quadratic in $a$.  It is going to be useful to
simplify the notation, so let us define
\begin{equation}
\begin{split}
  L^a=L_M^a\partial X^M 
  &= \sum(k_{(i)}-c_{G_i})\Tr(t_{a,R} g^{-1}\partial g)\ ,
  \\
  -R^a=-R_M^a\partial X^M 
  &= \sum(k_{(i)}-c_{G_i})\Tr(t_{a,L} \partial g g^{-1})\ ,
  \\
  M_{ab} &= \sum(k_{(i)}-c_{G_i})\Tr(t_{a,L} g t_{b,R} g^{-1}-t_{a,L} t_{b,L})\ ,
  \\
  \widetilde{M}_{ab} &=\sum(k_{(i)}-c_{G_i})\Tr(t_{b,L} g t_{a,R} g^{-1}
  -t_{a,R} t_{b,R})
  = M_{ba} + 2 H_{ab}\ ,
  \\
  G_{ab} &= \sum(k_{(i)}-c_{G_i})\lambda_i \half
  \Tr(t_{a,L} t_{b,L}+t_{a,R}t_{b,R})
  \\&=\sum(c_{G_i}-c_H)\half\Tr(t_{a,L} t_{b,L}+t_{a,R}t_{b,R})\ ,
  \\
  H_{ab} &= \sum(k_{(i)}-c_{G_i}) \half
  \Tr(t_{a,L} t_{b,L}-t_{a,R}t_{b,R})\ ,
  \\
  {\rm g}=g_{MN} \partial X^M\partial X^N
  &= \sum(k_{(i)}-c_{G_i}) \Tr(g^{-1}\partial g g^{-1}\partial g)\ .
\end{split}
\end{equation}
In this notation the action can be written as:
\begin{equation}
\begin{split}
  S_0^{\rm eff} = -\frac{1}{4\pi}\int d\tau \Bigl\{ &
  {\rm g} - 2a_\bz^a R_a - 2a_z^a L_a - 2a_\bz^a a_z^b(M_{ab}-G_{ab}+H_{ab})
  \\ &
  -a_z^a a_z^b(G_{ab}+H_{ab}) - a_\bz^a a_\bz^b(G_{ab}-H_{ab})
  \Bigr\}\ .
\end{split}
\end{equation}
Defining
\begin{equation}
\begin{split}
  &z^i = \left(\begin{array}{c} a_\bz^a \\ a_z^b \end{array} \right)\ ,
  \qquad
  B_i = \left(\begin{array}{c} R_a \\ L_b \end{array} \right)^T\ ,
  \\
  &A_{ij} = \left(\begin{array}{cc}
      G-H & M - (G-H) \\ M^T-(G-H)^T & G+H 
  \end{array}\right)
   =\left(\begin{array}{cc}
      G_- & M - G_- \\ \tilde{M}-G_+ & G_+ 
  \end{array}\right)\ ,
\end{split}
\end{equation}
where $G_+=G+H$ and $G_-=G-H$,
the action can be further simplified to
\begin{equation}
  S_0^{\rm eff} = -\frac{1}{4\pi}\int d\tau \Bigl\{
    {\rm g} - 2B_i z^i - z^i A_{ij} z^j
  \Bigr\}\ .
\end{equation}
Now we can complete the square, and get
\begin{equation}
  S_0^{\rm eff} = -\frac{1}{4\pi}\int d\tau \Bigl\{
    {\rm g} - A_{ij}(z+A^{-1}B)^i(z+A^{-1}B)^j + A^{kl}B_k B_l
  \Bigr\}\ ,
\end{equation}
where $A^{kl}\equiv(A^{-1})_{kl}$.

The equations of motion for $z$ ({\it i.e.,} the equations of motion for the
gauge fields $a_z$ and $a_\bz$) are now easily read off,
\begin{equation}
  \delta z \Rightarrow \qquad z^i = -A^{ik}B_k\ .
\end{equation}
Inserting this into the action, we end up with
\begin{equation}
  S^{\rm eff}_{\rm min} = -\frac{1}{4\pi}\int d\tau \Bigl[
  {\rm g} + B_k A^{kl} B_l \Bigr]\ .
\end{equation}
To write out this explicitly we need to invert the matrix $A_{ij}$. If
we write this inverted matrix as
\begin{equation}
  A^{-1} = \left(\begin{array}{cc} p & q \\ r & s \end{array}\right),
\end{equation}
then we can write
\begin{align}
  a_\bz^a &= -p_{ab}R_b - q_{ab}L_b\ ,
  \\
  a_z^a &= -r_{ab}R_b - s_{ab}L_b\ ,
\end{align}
and
\begin{equation}
\begin{split}
  S^{\rm eff}_{\rm min} &= -\frac{1}{4\pi}\int d\tau \Bigl[
    {\rm g} + R^a p_{ab} R^b + R^a (q_{ab}+r_{ba})L^b + L^a s_{ab} L^b
  \Bigr]
  \\
  &=-\frac{1}{4\pi}\int d\tau \Bigl[
    {\rm g}_{MN} +  R_M^a p_{ab} R_N^b 
    + R_M^a (q_{ab}+r_{ba})L_N^b + L_M^a s_{ab} L_N^b
  \Bigr]\partial X^M\partial X^N
  \\
  &=-\frac{1}{4\pi}\int d\tau ~\half C_{MN}\partial X^M\partial X^N\ .
\end{split}
\end{equation}
So, finding the coefficients  $C_{MN}$ means finding the matrices
$p,q,r,s$.
Explicitly,
\begin{equation}
  C_{MN}=2[ {\rm g}_{MN}
    + R_{M}^a p_{ab} R_{N}^b 
    + R_{M}^a (q_{ab}+r_{ba})L_{N}^b 
    + L_{M}^a s_{ab} L_{N}^b]\ .
\end{equation}
Note that $C_{MN}$ is not automatically symmetric.

Now let us recall the parameterization of the gauge groups.  The
generators of the gauge group $H=U(1)_A\times U(1)_B$, when acting on
the $H\subset SL(2,\mathbb{R})$ part are:
\begin{equation}
  t^{(1)}_{A,L} = \half\sigma_3\ ,
  \qquad t^{(1)}_{B,L} = 0\ ,
  \qquad t^{(1)}_{A,R} = -\frac{\delta}{2}\sigma_3\ ,
  \qquad t^{(1)}_{B,R} = -\frac{\lambda}{2}\sigma_3\ .
\end{equation}
The generators of $H$ when acting on the $H \subset SU(2)$
part are:
\begin{equation}
  t^{(2)}_{A,L} = 0\ , 
  \qquad t^{(2)}_{B,L} = 0\ , 
  \qquad t^{(2)}_{A,R} = 0\ , 
  \qquad t^{(2)}_{B,R}=-\frac{i}{2}\sigma_3\ .
\end{equation}
We note once more that this gauging leaves the global $SU(2)_L$
symmetry untouched, and so it will survive as a global symmetry of the
final model; the $SU(2)$ invariance of Taub--NUT. Finally, introduce
the generators of $H$ when acting on the fermionic part, $H\subset
SO(4)$:
\begin{equation}
\begin{array}{ll}
  t^{(f)}_{A,L} = \frac{1}{\sqrt{2}}\left(\begin{array}{cccc}
          0 & -Q_A & &\\ Q_A & 0 &&
          \\&& 0 & P_A \\ && -P_A & 0
        \end{array}\right)\ ,
  &
  t^{(f)}_{A,R} = -\frac{1}{\sqrt{2}}\left(\begin{array}{cccc}
          0 & -\delta & &\\ \delta & 0 &&
          \\&& 0 & 0 \\ && 0 & 0
        \end{array}\right)\ ,
  \\ \\
  t^{(f)}_{B,L} = \frac{1}{\sqrt{2}}\left(\begin{array}{cccc}
          0 & -Q_B & &\\ Q_B & 0 &&
          \\&& 0 & P_B \\ && -P_B & 0
        \end{array}\right)\ ,
  &
  t^{(f)}_{B,R} = -\frac{1}{\sqrt{2}}\left(\begin{array}{cccc}
          0 & -\lambda & &\\ \lambda & 0 &&
          \\&& 0 & 1 \\ && -1 & 0
        \end{array}\right)\ .
\end{array}
\end{equation}
Note that the $t_R$ are fixed by $(0,1)$ world--sheet supersymmetry,
while in the $t_L$, the $Q_{A,B}$ and $P_{A,B}$ are chosen to cancel
the anomaly {\it via} equation~\reef{eq:anomalies}.  The group
elements are chosen as:
\begin{align} g_1 &=
  = \frac{1}{\sqrt{2}}\left( \begin{array}{cc} e^\frac{t_+}{2}(x^2+1)^{1/2} &
      e^\frac{t_-}{2}(x^2-1)^{1/2}
      \\
      e^{-\frac{t_-}{2}}(x^2-1)^{1/2}&
      e^{-\frac{t_+}{2}}(x^2+1)^{1/2}
         \end{array}\right) \in SL(2,\mathbb{R})\ ,
  \\
  g_2 &=
  e^{\frac{i\phi}{2}\sigma_3}e^{\frac{i\theta}{2}\sigma_2}e^{\frac{i\psi}{2}\sigma_3} \\
  &=
  \left( \begin{array}{cc} 
          e^{\frac{i\phi_+}{2}}\cos\frac{\theta}{2} 
          & e^{\frac{i\phi_-}{2}}\sin\frac{\theta}{2}  
          \\
          -e^{-\frac{i\phi_-}{2}}\sin\frac{\theta}{2} 
          & e^{-\frac{i\phi_+}{2}}\cos\frac{\theta}{2}  
        \end{array}\right) \in SU(2)\ ,
  \\
  g_f &= \exp\Biggl\{\left( \begin{array}{cc} 
        \Phi_1 \frac{i\sigma_2}{\sqrt{2}} & 
           \\ & -\Phi_2 \frac{i\sigma_2}{\sqrt{2}}
        \end{array} \right)\Biggr\}
    = \left( \begin{array}{cccc}
          \cos\frac{\Phi_1}{\sqrt{2}} & \sin\frac{\Phi_1}{\sqrt{2}} && \\
          -\sin\frac{\Phi_1}{\sqrt{2}} & \cos\frac{\Phi_1}{\sqrt{2}} &&\\
          && \cos\frac{\Phi_2}{\sqrt{2}} & -\sin\frac{\Phi_2}{\sqrt{2}} \\
          && \sin\frac{\Phi_2}{\sqrt{2}} & \cos\frac{\Phi_2}{\sqrt{2}}
        \end{array} \right) \in SO(4)\ ,
\end{align}
where $t_L,t_R,x\in \mathbb{R}$, $\theta\in(0,\pi)$,
$\phi\in(0,2\pi)$, $\psi\in(0,4\pi)$, and $\Phi_1$ and $\Phi_2$ are
$2\pi$ periodic.  Also, $\phi_\pm=\phi\pm\psi$ and $t_\pm=t_L\pm t_R$.
We have already gauge--fixed the fermionic sector.

To find the coefficients $C_{MN}$ we now have to compute the group
manifold metric $g_{MN}$ and the vectors $L_M$ and $R_M$. We also have
to compute the matrix $A_{ij}$ and find its inverse. This is all
relatively straightforward and the details, involving a number of
rather messy expressions, are left out.  Having completed this task,
we must worry about the effects of re--fermionization.

\subsubsection{Re--fermionization and Back Reaction on Metric}
Assume that the local part of the action can be written (where we have
re-introduced dependence on worldsheet space as well as time, which is
necessary to deduce the $B$--field)
\begin{equation}
   S = \half\int d^2\!z\,
     C_{MN}\partial X^M\bar{\partial}X^N\ .
\end{equation}
This expression can be rewritten as follows:
\begin{align}
  S &= \half\int d^2\!z\,
    C_{MN}\partial X^M\bar{\partial}X^N
  \\ \nonumber
  &= \half\int d^2\!z \Bigr[
    C_{\mu\nu}\partial X^\mu\bar{\partial}X^\nu
    +A^i_\mu(\partial X^\mu\bar{\partial}\Phi^i
      +\bar{\partial}X^\mu\partial\Phi^i)
    +B^i_\mu(\partial X^\mu\bar{\partial}\Phi^i
      -\bar{\partial}X^\mu\partial\Phi^i)
  \\ &\qquad\qquad\quad
    +R_{ij}\half(\partial\Phi^i\bar{\partial}\Phi^j
    +\bar{\partial}\Phi^i\partial\Phi^j)
    +F_{ij}\half(\partial\Phi^i\bar{\partial}\Phi^j
    -\bar{\partial}\Phi^i\partial\Phi^j) \Bigr]\ ,
  \\ \nonumber
   &= \half\int d^2\!z \Bigl[
  (C_{\mu\nu}-R^{ij}A^i_\mu A^j_\nu) 
  \partial X^\mu\bar{\partial}X^\nu
  +R_{ij}(\partial\Phi^i+R^{ik}A^k_\mu\partial X^\mu)
  (\bar{\partial}\Phi^j+R^{jl}A^l_\nu\bar{\partial}X^\nu)
  \\
  &\qquad\qquad\qquad B^i_\mu(\partial X^\mu\bar{\partial}\Phi^i
  -\bar{\partial}X^\mu\partial\Phi^i)
  +F_{ij}\half(\partial\Phi^i\bar{\partial}\Phi^j
  -\bar{\partial}\Phi^i\partial\Phi^j) \Bigr]\ ,
\label{referm}
\end{align}
where
\begin{equation}
\begin{array}{ll }
  A^i_\mu = C_{\mu i}+C_{i\mu}\ ,
  &
  B^i_\mu = C_{\mu i}-C_{i\mu}\ ,
  \\
  R_{ij} = C_{(ij)}\ ,
  &
  F_{ij} = C_{[ij]}\ .
\end{array}
\end{equation}
Note that in the zero mode sector where we keep only symmetric terms,
which means $F_{ij}=0$ and $B^i_\mu=0$.  This is (almost) the form
required for re--fermionization, and we can read off the metric from
the first term. 
Before refermionisation, we must rescale the $\Phi$s in the action
\reef{referm} that the term $R_{ij}\partial\Phi^i{\bar
  \partial}\Phi^j$ becomes $\delta_{ij}\partial{\widetilde\Phi}^i{\bar
  \partial}{\widetilde\Phi}^j$. This is done by:
\begin{equation}
\Phi^i=U^i_j{\widetilde \Phi}^j\ ,
  \label{eq:rescale}
\end{equation}
with $R_{ij}U^i_kU^j_l=\delta_{kl}$. This corrects the $A^i_\mu$ to
${\cal A}^i_\mu = R^{ij}A^j_\mu$, where $R^{ij}=(R^{-1})_{ij}$. The
spacetime metric is then:
\begin{align}
  G_{\mu\nu} &= C_{(\mu\nu)}-R^{ij}A^i_{(\mu}A^j_{\nu)}
  = G^0_{\mu\nu} - {\cal A}^i_{\mu}{\cal A}^j_{\nu}\ .
\end{align}

Carrying out the computation, we find that the final expression for
the exact metric simplifies in a remarkable way to the following
(using equation~\reef{eq:centralcharge} we write $k_1=k$, $k_2=k-4$):
\begin{equation}
\begin{split}
  ds^2 &= G_{\mu\nu}dX^\mu dX^\nu
  \\
  &= (k-2)\Bigl\{
  \frac{dx^2}{x^2-1} - \frac{x^2-1}{D(x)}  (dt+2\lambda A^M_\phi d\phi)^2
  + d\theta^2+\sin^2\!\theta d\phi^2 \Bigr\}\ ,
\end{split}
\end{equation}
where
\begin{equation}
  D(x) = (x+\delta)^2 -\frac{4}{k+2}(x^2-1)\ ,
  \label{eq:dee}
\end{equation}
and $2A^M_\phi=\pm1-\cos\theta$ is a Dirac monopole connection where
$\pm$ refers to the N(S) pole on the $S^2$. The $\pm1$ can be gauged
away by {\it e.g.,} a shift of $t$ to match the form given in
section~\ref{sec:introduction}. The dilaton is generated by the
effects of two Jacobians. One comes from the determinant, $\det A$,
arising from integrating out the gauge fields, but there is another
contribution coming from the change of variables from $\Phi$ to
${\widetilde\Phi}$.  That Jacobian is:
\begin{equation}
  \left|\frac{\partial\Phi}{\partial{\widetilde\Phi}}\right|={\rm
    det}U=({\rm det}R)^{-1/2}\ .
  \label{eq:jacob}
\end{equation}
This results in\cite{Buscher:1988qj}:
\begin{equation}
  e^{2\Phi} = (\det A)^{-\frac{1}{2}}(\det R)^{-\frac{1}{2}}\ ,
\end{equation}
where the determinants can be written as follows. Define
\begin{equation}
p=k-2+2P_B\ ,\quad q = (k+2)\delta+2Q_A\ ,\quad r = (k+2)\lambda+2Q_B\ .
  \label{eq:definitions}
\end{equation}
Then
\begin{equation}
\begin{split}
  \det A =\Delta(x)=\left((k-2)px-(2P_A r - pq)\right)^2+4(r^2-p^2)\ ,
\end{split}
\label{eq:deta}
\end{equation}
and
\begin{equation}
\det R=4(k+2)(k-2)^3\frac{D(x)}{\Delta(x)}\ .
  \label{eq:detr}
\end{equation}
The result is that the exact dilaton is:
\begin{equation}
\Phi-\Phi_0=-\frac{1}{4}\ln(D(x))\ ,
  \label{eq:exactdilaton}
\end{equation}
where we have absorbed a non--essential constant into the definition
of $\Phi_0$.  The expressions for the exact fields
$B_{\mu\nu},A^i_\mu$ are somewhat involved, but straightforward to
read off. We will not list them here, as we will not need them in what
follows.

As a useful check on our procedure, it is worth noting that the large
$k$ limit gives the expressions originally written in
ref.\cite{Johnson:1994jw}.  In this limit, we get $D\to (\delta+x)^2$,
and the metric becomes that given in
equation~\reef{eq:taubnutlowenergyextended}, and the dilaton becomes:
\begin{equation}
  \Phi-\Phi_0   \to -\frac{1}{2}\ln(x+\delta)\ .
\end{equation}

\subsection{Properties of the Exact Metric}
\label{sec:properties}
As already stated in the previous section, the final result for the
exact spacetime metric is (after a trivial shift in $t$):
\begin{eqnarray}
ds^2&=&(k-2)
\left( \frac{dx^2}{x^2-1}+F(x)(dt-\lambda\cos\theta d\phi)^2+d\theta^2+\sin^2\theta d\phi^2\right)\ ,
\nonumber\\
{\rm where}\quad F(x)&=&-\frac{x^2-1}{D(x)}=-\left(\frac{(x+\delta)^2}{x^2-1}-\frac{4}{k+2}\right)^{-1}\ .
  \label{eq:taubnutexactextended}
\end{eqnarray}
This is a pleasingly simple form to result from such an involved
computation. In fact, its relation to the leading order result is
reminiscent in form to the relation between the leading order and
exact results for the black hole $SL(2,\IR)/U(1)$
model\cite{Witten:1991yr,Dijkgraaf:1992ba}.

\begin{figure}[ht]
\begin{center}
  \includegraphics[scale=0.5]{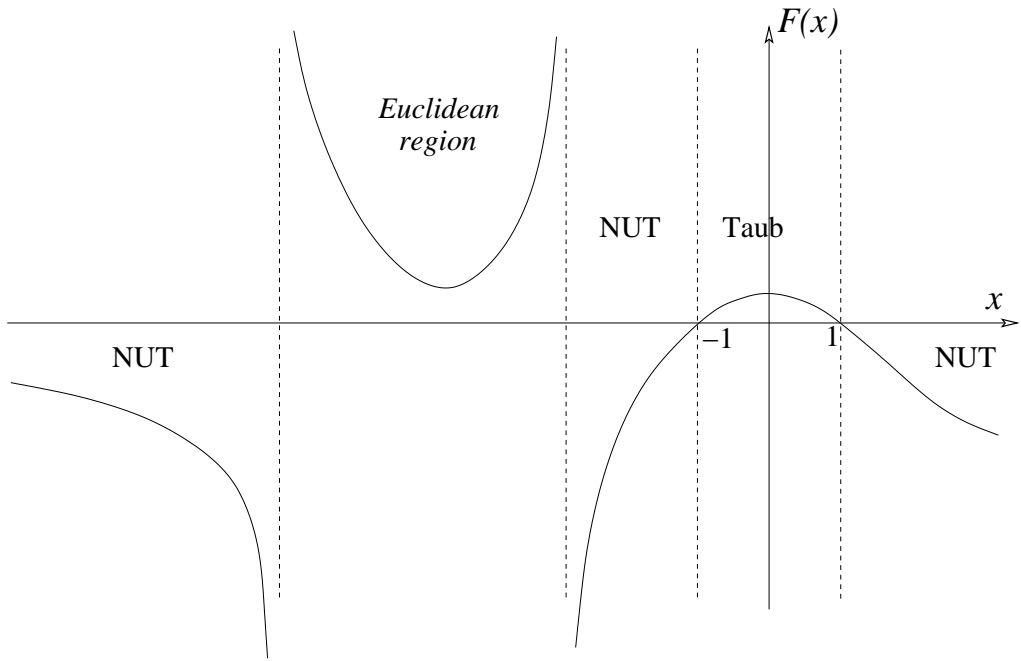}
\end{center}
\caption{\small The various regions in the stringy Taub--NUT geometry for arbitrary $k$, with all  $1/k$ corrections included. Compare to the leading order result in figure~\ref{zeros}. Note that the singularity splits in order to incorporate a finite sized region of Euclidean signature in the second NUT region.}
\label{zeros2}
\end{figure}

It is interesting to sketch the behaviour of $G_{tt}=F(x)$, as it
contains the answer to our original questions about the fate of the
Taub and NUT regions of the spacetime once the contributions of the
stringy physics are included. This result is plotted in
figure~\ref{zeros2}, and it should be contrasted with
figure~\ref{zeros}.

Several remarks are in order. The first is that the Taub and NUT
regions, although modified somewhat, survive to all orders. The second
is that the local structure of the chronology horizons separating
these regions is completely unaffected by the stringy corrections!
$F(x)$ still vanishes at $x=\pm1$ and furthermore for $x=1-\tau$ where
$\tau$ is small, the metric of the $(\tau,\xi)$ space (the space over
each point of the $S^2$) becomes:
\begin{equation}
ds^2=(k-2)\left(-(2\tau)^{-1}d\tau^2+\frac{2\tau}{(1+\delta)^2}d\xi^2\right)\ ,
  \label{eq:misnerlike2}
\end{equation}
which is again of Misner form.

\begin{figure}[ht]
\begin{center}
  \includegraphics[scale=0.5]{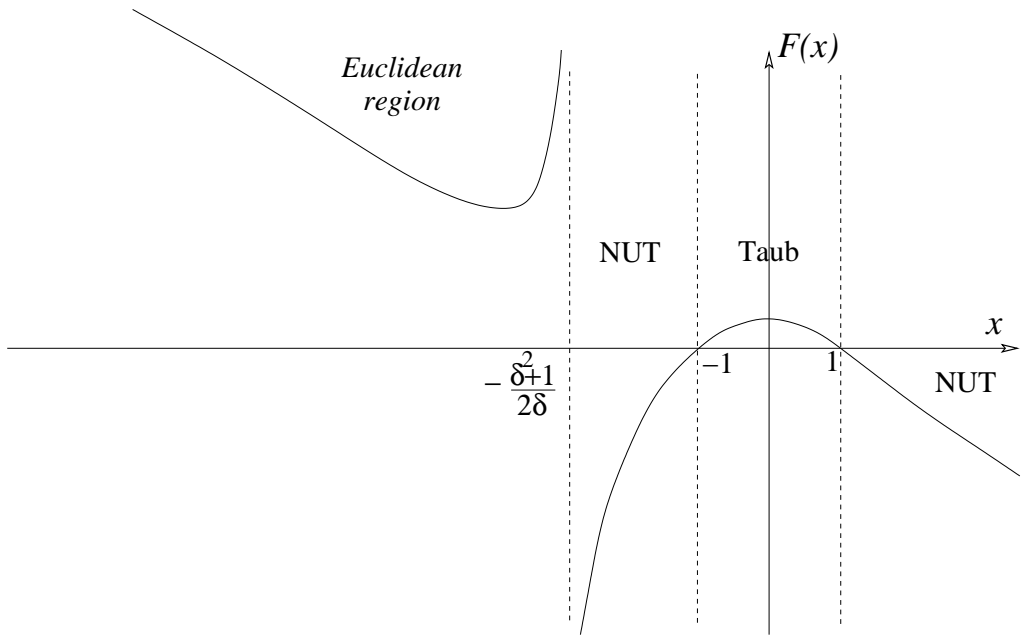}
\end{center}
\caption{\small The various regions in the stringy Taub--NUT geometry for the smallest value of $k$ possible.  This is the ``most stringy'' geometry. Compare to the leading order result in figure~\ref{zeros} and the intermediate $k$ result in figure~\ref{zeros2}. The Euclidean region has grown and occupied the entire region to the left, making the second nut region of finite extent.}
\label{zeros3}
\end{figure}

Notice that the singularity we observed in $F(x)$ (and the spacetime)
has now split into two.  Recalling the definition of $D(x)$ given in
equation~\reef{eq:dee}, we can write the Ricci scalar as:
\begin{equation}
\begin{split}
  R = -\frac{1}{2(k-2)D^2}\Bigl[& 2D(x^2-1)D^{\prime\prime}
  -3(x^2-1)\left(D^\prime\right)^2
  +\lambda^2(x^2-1)D+6xDD^\prime \Bigr]\ ,
\end{split}
\end{equation}
(where a prime means $d/dx$). $R$ diverges if and only if $D(x)=0$.
These singularities are located at:
\begin{equation}
x_{\pm}=\frac{-\delta\pm\sqrt{a^2+a(\delta^2-1)}}{(1-a)}\ ,\quad a=\frac{4}{k+2}\ ,
  \label{eq:singularlocations}
\end{equation}
and the region in between them has Euclidean signature. Such a region
was noticed in ref.\cite{Perry:1993ry} in the context of the exact
metric for the $SL(2,\IR)/U(1)$ coset giving the two dimensional black
hole. This region remains entirely within the second NUT region,
however, and never approaches the Misner horizons. Its size goes as
$1/(k-2)$. The model only seems to make sense for $k>2$, of course,
and it interesting to note that the limiting behaviour of this metric
as $k\to2^+$ is that the Euclidean region grows until it fills the
entire left hand side of the sketch (see figure~\ref{zeros3}), with
one singularity at $x=-(\delta^2+1)/(2\delta)$, and the other, when
last seen, was moving off to $x=-\infty$.

\section{Discussion}
\label{sec:discussion}
Our goal was to identify a stringy laboratory for the study of a
number of issues of interest, which allows a controlled study of
various physical phenomena. Closed time--like curves are very common
in General Relativity, but the theory is silent about their physical
role in a complete theory of gravity. They can appear after a
cosmology passes through a certain type of spacelike ``Big Crunch''
singularity, and it is natural to wonder if the full theory somehow
modifies the geometry in a way which obstructs this process of
formation, realizing the so--called chronology protection
conjecture\cite{Hawking:1992nk}.  The model upon which a great deal of
the study within General Relativity has been focused is the Taub--NUT
spacetime (or local parts of it).  Quite satisfyingly, this is
precisely the model that we study here, furthering earlier work which
showed how to embed it into string theory in a way which allows a
complete definition in terms of conformal field theory.

The study of the model we performed here was to go beyond the low
energy truncation and compute the all orders in $\alpha^\prime$
geometry, thereby including the effects of the entire string spectrum
on the background. Our embedding (into heterotic string theory) was
chosen so as to permit such corrections to occur, at least in
principle. Somewhat surprisingly (perhaps) we found that the key
features of the Taub--NUT geometry persist to all orders. This
includes the fact that the volume of the universe in the Taub
cosmology vanishes as a circle shrinks to zero size, at the junction
(described by Misner space) where the CTCs first appear. There is no
disconnection of the Taub region from the NUT regions containing the
CTCs, to all orders in $\alpha^\prime$. Note that the strength of the
string coupling near the junctions is not particularly remarkable, and
so an appeal to severe corrections purely due to string loops may not
help modify the geometry further.

We have therefore ruled out a large class of possible modification to
the geometry which could have destroyed the chronology horizons and
prevented the formation of the CTC regions (from the point of view of
someonestarting in the cosmological Taub region). As remarked upon in
the introduction, there is still the possibility that there is an
instability of the {\it full} geometry to backreaction by probe
particles or strings. A large class of such effects are likely missed
by our all orders computation of the metric. There are studies of
Misner space in various dimensions (in its orbifold representations)
that signal such an
instability\cite{Lawrence:2002aj,Horowitz:2002mw,Liu:2002kb}, and the
fate of the chronology horizons embedded in our geometry should be
examined in the light of those studies. The nature of the spacetime in
which they are embedded is important, however, and so it seems that
the relevant geometry to study such backreaction effects is the fully
corrected geometry we have derived here, since it takes into account
the full $\alpha^\prime$ effects.

Quantum effects may well be important even though the string coupling
is not strong at the chronology horizons, and even if there are no (as
we have seen here) modifications due to $\alpha^\prime$ corrections.
Radically new physics can happen if there are the right
sort of special (for example, massless) states arising in the theory
there together with (crucially) certain types of new physics. Strings
wrapped on the $t$--circle are candidate such states.  Following these
states could shed new light on the validity of the geometry if they
are accompanied by the appropriate physics, such as in the mechanism
of ref.\cite{Johnson:1999qt}. Such probe heterotic strings are hard to
study in the sigma model approach, but it would be interesting to
undergo such an investigation. The study of probes directly in the
full conformal field theory (\ie, without direct reference to the
geometry) may well be the most efficient way to proceed.

Another (less often considered) possibility is that the result of this
paper is a sign that the theory is telling us that it is perfectly
well--defined in this geometry. The conformal field theory is (at face
value) well--defined, and there are no obvious signs of a pathology.
Perhaps string theory is able to make sense of all of the features of
Taub--NUT. For example, the shrinking of the spatial circle away to
zero size at the Big Bang or Big Crunch might not produce a pathology
of the conformal field theory even through there might be massless
states appearing from wrapped heterotic strings. They might simply be
incorporated into the physics in a way that does not invalidate the
geometry: The physics, as defined by the world--sheet model, would
then carry on perfectly sensibly through that region.  This would mean
that would be another geometry that a dual heterotic string sees which
is perfectly smooth through this region. It would be interesting to
construct this geometry\footnote{The right--handed world--sheet parity
  flip which generates a dual geometry is no longer achievable by
   axial--vector duality as in simpler cases such as the
  $SL(2,\IR)/U(1)$ black hole\cite{Dijkgraaf:1992ba,Kiritsis:1991zt}.
  It only works for $\delta=\pm1$, $\lambda=0$. Here, it is natural
  to explore whether $\delta\to-\delta$ combined with other actions
  might generate it, but a fibre--wise duality rather like that which
  relates\cite{Ooguri:1996wj,Tong:2002rq} an NS5--brane to an ALE
  space might be more appropriate.}.

In this scenario, if we accept that the conformal field theory is
telling us that the stringy physics is well behaved as it goes through
from the Taub region to the NUT region, we have to face the
possibility that the CTCs contained in the NUT regions might well be
acceptable, and part of the full physics as well.

While it is perhaps too early to conclude this with certainty, it is
worth noting that most objections that are raised about physics with
CTCs are usually ones based on paradoxes arrived at using macroscopic
and manifestly classical reasoning, or reasoning based on our very
limited understanding of quantum theory outside of situations where
there is an asymptotic spacetime region to which we make reference.
Some CTCs fall outside of those realms, opening up new possibilities.
We must recall that time, just like space, is supposed to arrive in
our physics as an approximate object, having a more fundamental
quantum mechanical description in our theory of quantum gravity. The
ubiquity of CTCs in theories of gravity might be a sign that
(appropriately attended to) they are no more harmful than closed
spatial circles.  Rather than try to discard CTCs, we might also keep
in mind the possibility that they might play a natural role in the
full theory, when we properly include quantum mechanics.  Here, we saw
them remain naturally adjoined to a toy cosmology, surviving all
$\alpha^\prime$ corrections.  This is just the sort of scenario where
CTCs might play a role in Nature: A natural way to render meaningless
the usual questions about the lifetime of the universe prior to the
``Big Bang'' is to have the Big Bang phase adjoined to a region with
CTCs\footnote{Although it is in the very different context of eternal
  inflation, the role of CTCs in cosmology has been speculated about
  before\cite{Gott:1998pm}.}.  This is an amusing alternative to the
usual scenarios, and may be naturally realized within string theory,
or its fully non--perturbative successor.

\section*{Acknowledgments}
HGS is supported by the Research Council of Norway, and by an ORS
award at Durham. HGS thanks the Physics Department at USC for support
and hospitality. CVJ is grateful to Andrew Chamblin for mentioning (at
the String Cosmology workshop at the ITP at UCSB in November) that
Taub--NUT was of interest to Relativists in the study of CTCs and for
pointing out ref.\cite{Hiscock:1982vq}. CVJ thanks the group and
visitors at Caltech for interesting comments and remarks during an
enjoyable seminar in February, particularly Mike Douglas, Jaume
Gomis and Hirosi Ooguri. CVJ also thanks Itzhak Bars, Eric Gimon and
Petr Ho\u rava for interesting conversations.


\providecommand{\href}[2]{#2}\begingroup\raggedright\endgroup

\end{document}